\documentclass[twocolumn,english,aps,reprint,groupedaddress,notitlepage,nobibnotes,nofootinbib,preprintnumbers,showpacs]{revtex4-1}
\pdfoutput=1
\usepackage{lmodern}

\usepackage[T1]{fontenc}
\usepackage[latin9]{inputenc}
\usepackage{geometry}
\usepackage{color}
\usepackage{babel}
\usepackage{mathtools}
\usepackage{amsmath}
\usepackage{amssymb}
\usepackage{graphicx}
\usepackage{esint}
\usepackage{array}
\usepackage[unicode=true,pdfusetitle,
 bookmarks=true,bookmarksnumbered=false,bookmarksopen=false,
 breaklinks=false,pdfborder={0 0 1},backref=false,colorlinks=true]
 {hyperref}
\hypersetup{
 citecolor=black,linkcolor=black,urlcolor=black}
\makeatletter

 \@ifundefined{textcolor}{}
 {
   \definecolor{BLACK}{gray}{0}
   \definecolor{WHITE}{gray}{1}
   \definecolor{RED}{rgb}{1,0,0}
   \definecolor{GREEN}{rgb}{0,1,0}
   \definecolor{BLUE}{rgb}{0,0,1}
   \definecolor{CYAN}{cmyk}{1,0,0,0}
   \definecolor{MAGENTA}{cmyk}{0,1,0,0}
   \definecolor{YELLOW}{cmyk}{0,0,1,0}
 }
\usepackage{babel}
\makeatletter
\def\simgt{\mathrel{\lower2.5pt\vbox{\lineskip=0pt\baselineskip=0pt
           \hbox{$>$}\hbox{$\sim$}}}}
\def\simlt{\mathrel{\lower2.5pt\vbox{\lineskip=0pt\baselineskip=0pt
           \hbox{$<$}\hbox{$\sim$}}}}
\makeatother

\newcommand{\be}{\begin{equation}}
\newcommand{\ee}{\end{equation}}
\newcommand{\bea}{\begin{eqnarray}}
\newcommand{\eea}{\end{eqnarray}}

\newcommand{\fig}[1]{Fig.~\ref{#1}}

\newcommand{\Eqn}[1]{Eq.~\eqref{#1}}
\newcommand{\eqn}[1]{Eq.~\eqref{#1}}

\newcommand{\vev}[1]{\langle #1 \rangle}
\newcommand{\bra}[1]{\langle #1 |}
\newcommand{\ket}[1]{| #1 \rangle}

\def\NeqFour{{\cal N} = 4}
\def\spa#1.#2{\left\langle#1\,#2\right\rangle}
\def\spb#1.#2{\left[#1\,#2\right]}
\def\Section#1{\vskip.2 cm \noindent{\it #1.}}

\begin{document}

\preprint{\hbox{UCLA/18/TEP/103 } }

\title{Dual Conformal Structure Beyond the Planar Limit}

\author{Zvi~Bern, Michael Enciso, Chia-Hsien Shen and Mao Zeng}
\affiliation{Mani L. Bhaumik Institute for Theoretical Physics\\
	Department of Physics and Astronomy\\
	University of California at Los Angeles\\
	Los Angeles, CA 90095, USA\\}


\begin{abstract}
The planar scattering amplitudes of $\NeqFour$ super-Yang--Mills
theory display symmetries and structures which underlie their
relatively simple analytic properties such as having only logarithmic
singularities and no poles at infinity.  Recent work shows in various
nontrivial examples that the simple analytic properties of the planar
sector survive into the nonplanar sector, but this has yet to be
understood from underlying symmetries.  Here we explicitly show that
for an infinite class of nonplanar integrals that covers all
subleading-color contributions to the two-loop four- and five-point
amplitudes of $\NeqFour$ super-Yang--Mills theory, symmetries
analogous to dual conformal invariance exist.  A natural conjecture is
that this continues to all amplitudes of the theory at any loop order.
\end{abstract}
\maketitle

\Section{Introduction}
Recent years have seen significant advances in constructing scattering
amplitudes, especially for planar $\NeqFour$ super-Yang--Mills (sYM)
theory.  A key feature of planar $\NeqFour$ sYM theory that makes this
progress possible is its remarkable symmetries and structures.  These
include dual conformal symmetry~\cite{DualConformal}, Yangian
symmetry~\cite{Yangian}, integrability~\cite{BeisertStaudacher}, a
dual interpretation of scattering amplitudes in terms of Wilson
loops~\cite{WilsonLoops}, uniform
transcendentality~\cite{UniformTransc}, structures that aid various
bootstraps~\cite{BassoOPE,DixonBootstrap}, and even an all-loop
resummation of four- and five-point amplitudes~\cite{BDS}.  Scattering
amplitudes have been reformulated using on-shell diagrams and the
positive Grassmannian~\cite{Grassmannian}, which culminated in the
geometric concept of the amplituhedron~\cite{Amplituhedron}.  Some of
these advances have been helpful in quantum chromodynamics relevant
for collider physics, including improved ways for dealing with
polylogarithms that arise in multiloop computations~\cite{Symbols} and
for finding good
choices~\cite{HennIntegrals,nonplanarDlog,nonplanarAmplituhedron}
of integral bases that simplify their evaluation.  In fact, the
integrals we analyze here for the two-loop five-point
amplitude~\cite{JJHenrikFivePointTwoLoopN4,nonplanarAmplituhedron} are
useful choices for the basis of master integrals for $2$-to-$3$
scattering in generic theories~\cite{HennFivePtTwoloop}.

These symmetries and structures impose nontrivial constraints on the
analytic properties of planar $\NeqFour$ sYM amplitudes.
In particular, the loop-level color-ordered amplitudes $\mathcal{M}_{123\ldots n}$ can be
written as
\begin{equation}
\mathcal{M}_{123\ldots n} = \text{PT}_{123 \ldots n}\int \mathcal{I} \,, 
\label{eq:Amplituhedron}
\end{equation}
where the integrand $\mathcal{I}$ has only logarithmic singularities, no poles at
infinity~\cite{Grassmannian}, and unit leading singularities
\cite{NimaAllLoopIntegrand} as tied to the amplituhedron~\cite{Amplituhedron}. The prefactor
$\text{PT}_{123\ldots n}$ is the standard Parke-Taylor factor~\cite{ParkeTaylor},
as defined in e.g.~Ref.~\cite{nonplanarAmplituhedron}.

It is unclear how to define dual conformal symmetry in the nonplanar
sector given the lack of dual variables to define the symmetry.
However, as shown in a variety of
examples~\cite{Nimanonplanar,nonplanarDlog,nonplanarAmplituhedron}, the
key analytic properties of the planar sector implied by its symmetries
carry over to the nonplanar sector, even if the symmetries are
unclear.  In each example, the full amplitude can be expressed
as~\cite{nonplanarParkeTaylor}
\begin{equation}
\mathcal{M} = \sum_{k, \sigma, j} a_{\sigma, k, j} c_k \text{PT}_{\sigma} \int \mathcal{I}^j \,,
\label{eq:npAmplituhedron}
\end{equation} 
where the $a_{\sigma, k, j}$ are rational numbers, the $c_k$ are color
factors, the $\text{PT}_{\sigma}$ are the Parke-Taylor factors
corresponding to an ordering $\sigma$ of external particles, and the
$\mathcal{I}^j$ are integrands with only logarithmic singularities, no
poles at infinity, and unit leading singularities.
\Eqn{eq:npAmplituhedron} is a natural extension of
\eqn{eq:Amplituhedron} to the nonplanar sector.  It is nontrivial
that such a representation exists where each integrand is
expressed in terms of local diagrams. Some structures of the non-planar sector were also explored
at the level of on-shell diagrams~\cite{nonPlanarOnShellDiagrams,NonplanarYangian}.

In the present paper we address the following question: Can we
identify a hidden symmetry associated with the simple analytic
properties for the nonplanar sector uncovered in
Refs.~\cite{Nimanonplanar,nonplanarDlog, nonplanarAmplituhedron}?
Building on the initial studies in Ref.~\cite{EarliernonplanarSym}, we
answer this question affirmatively and demonstrate that the integrands
$\mathcal{I}^j$ in (\ref{eq:npAmplituhedron}) encoding the simple
analytic structure of the full two-loop four- and five-point
amplitudes all have hidden symmetries related to dual conformal
invariance.  These are not hidden symmetries of the full amplitude,
but of individual components of the amplitudes, analogous to the
situation with dual conformal symmetry in the planar case
\eqref{eq:Amplituhedron}.  We also identify an infinite class of
nonplanar integrands with the hidden symmetry.  In many cases these
symmetries rely on nontrivial identities, making it all the more
striking that a symmetry actually exists.

\begin{figure}
\includegraphics[scale=.1]{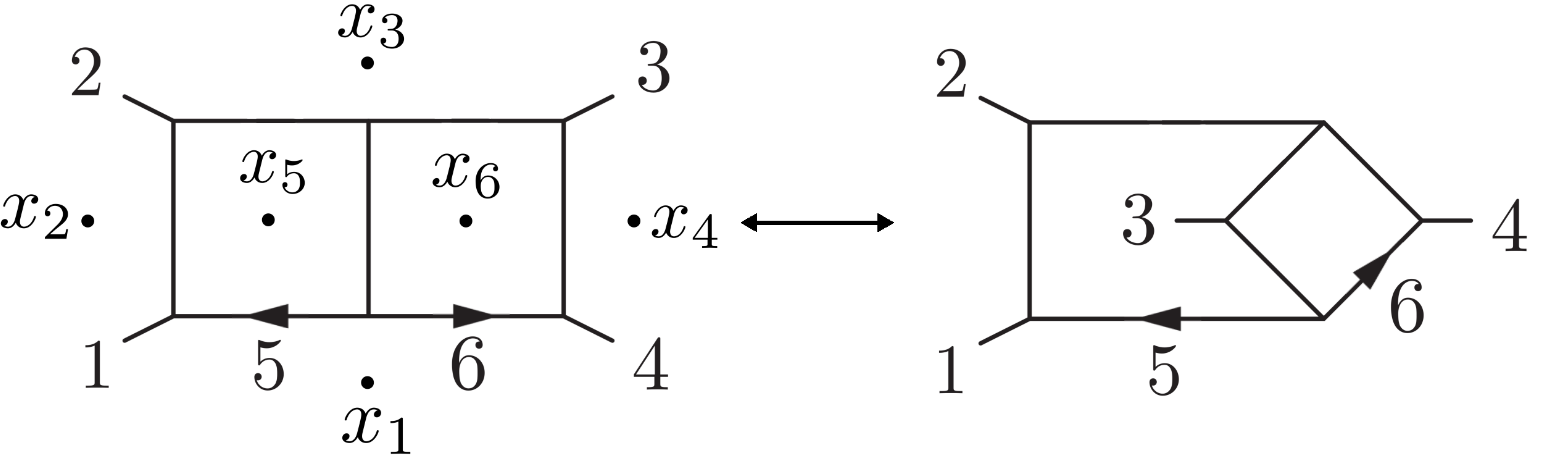}
\caption{Planar double box with dual coordinates and the crossed-box related to it by moving leg 3 to the central rung.}
\label{fig:fourPoints}
\end{figure}

\Section{Dual coordinates and conformal symmetry}
To set up our discussion of hidden symmetries in the nonplanar sector,
we first briefly review dual conformal symmetry in the planar
sector~\cite{DualConformal}.  In general, the momenta (corresponding to edges or lines) in
\emph{any} planar diagram can be represented as the difference of
adjacent dual coordinates (corresponding to regions).  For example, the momenta 
in the planar double-box diagram on the left of Figure~\ref{fig:fourPoints}
can be expressed as
\begin{align}
p_1 = x_2 - x_1\,, \hskip .5 cm &
p_2 = x_3 - x_2\,,  \hskip .1 cm &
p_3 = x_4 - x_3\,, \nonumber \\ 
p_4 = x_1 - x_4\,, \hskip .5 cm &
l_5 = x_5-x_1 \,,  \hskip .1 cm &
l_6 = x_1-x_6 \,, \label{TwoLoopDualVars}
\end{align}
where the $p_i$ are external momenta, $l_5$ and $l_6$ are the loop
momenta, and $x_i$ are the dual coordinates with all Lorentz indices
omitted. We can perform infinitesimal conformal transformations on
these dual coordinates,
\begin{equation}
\delta x_i^{\mu} = \frac{1}{2}x_i^2 b^{\mu} - (x_i\cdot b)x_i^{\mu} \,,
\label{eq:dual_tranform}
\end{equation}
where $b^{\mu}$ is an infinitesimal boost vector. The transformation of the square of proper distance is
\begin{equation}
\delta (x_i-x_j)^2 = -b \cdot (x_i+x_j)\,(x_i-x_j)^2\,.
\end{equation}
In general, if a quantity $f$ transforms as $\delta f = w f$ with $w$
a local function, we say $f$ \emph{rescales} under the transformation
with weight $w$.  Thus, under dual conformal transformations,
$(x_i-x_j)^2$ carries a weight $-b \cdot (x_i+x_j)$.  Note that all
massless external legs remain on-shell after the transformation.
All the inverse propagators have the form
$(x_i-x_j)^2$. This implies that locality is maintained for planar
loop integrals under dual conformal transformations and allows us to
construct simple functions that are invariant.

As a simple illustration, consider an integral associated with the
planar double box,
\begin{equation}
I = \int d^Dx_5 d^Dx_6 \frac{s^2 t}{\prod_k \rho_k}\,,
\label{DoubleBox}
\end{equation}
where $s=(x_1 - x_3)^2 = (p_1+p_2)^2$ and $t= (x_2 - x_4)^2 = (p_2+p_3)^2$. 
The inverse Feynman propagators $\rho_k$ in dual coordinates are
\begin{align}
\rho_1 =(x_5-x_1)^2\,, & \hskip .3 cm \rho_2 = (x_5-x_2)^2 \,, & \rho_3 = (x_5-x_3)^2\,, \nonumber\\
\rho_4= (x_5-x_6)^2\,, & \hskip .3 cm \rho_5 = (x_6-x_1)^2 \,, & \rho_6 = (x_6-x_4)^2\,, \nonumber\\
\rho_7 = (x_6-x_3)^2\,.
\end{align}
In what follows, we will be interested in the integrand $\mathcal{I}$, defined by $I = \int\mathcal{I}$.
With this numerator the integrand has a
hidden symmetry exposed by using the dual
variables~\cite{DualConformal}.  Performing the dual conformal
transformation on the integrand \eqref{DoubleBox} yields
\begin{equation}
\delta \mathcal{I} = -(D-4) \big(b\cdot (x_5+x_6)\big)\mathcal{I} \,,
\end{equation}
where we used
\begin{equation}
\delta (d^Dx_i) = 
\Big(\frac{\partial \delta x_i^{\mu}}{\partial x_i^{\mu}}\Big) d^Dx_i = -D(b\cdot x_i)\,d^Dx_i \,. 
\label{eq:deltaMeasure}
\end{equation}
For $D=4$ space-time dimensions this integrand is invariant under dual
conformal transformations, which is what motivated the choice of
numerator $s^2 t$.  Outside $D=4$, this is reminiscent of $\epsilon$-form differential
equations~\cite{HennIntegrals}, but without doubled propagators on the
right hand side before reduction to a basis~\cite{DENoDoubledProp,
  EarliernonplanarSym}.

What is the relevance of this symmetry? It turns out that \emph{all}
integrands of planar $\NeqFour$ sYM amplitudes possess this property,
which then leads to nontrivial constraints on the amplitude after
integration. This is the celebrated \emph{dual conformal
  symmetry}~\cite{DualConformal} which has spurred many developments. In
the following we identify an analogous symmetry in a class of
nonplanar diagrams.

\begin{figure}[tb]
\includegraphics[scale=.076]{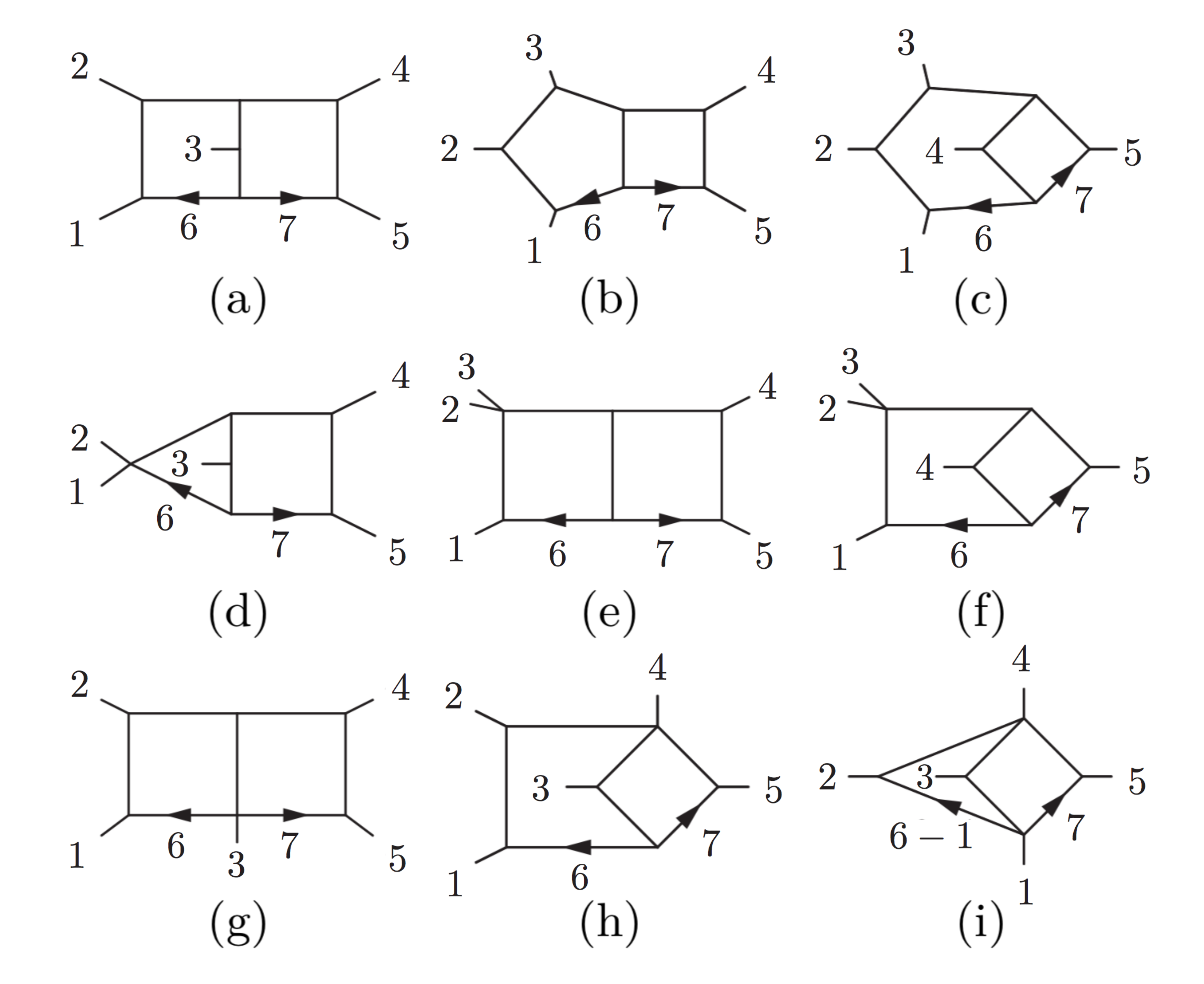}
\vskip -.4 cm 
\caption{Diagrams (a)-(i) compose the two-loop five-point
  amplitude in Ref.~\cite{nonplanarAmplituhedron}. 
}
\label{fig:fivePointAllDiags}
\end{figure}

\Section{Nonplanar extension} While there are no known global variables for
generic nonplanar diagrams, it is natural to require that, as for the
planar case, a non-planar analog of dual conformal transformations
also maintains the local structure for inverse propagators, $\delta
\rho_k \propto \rho_k$~\cite{EarliernonplanarSym}.  We start by
considering a nonplanar diagram that can be made planar by moving the
location of one external leg carrying momentum $p_k^{\mu}$.  This is an infinite class
of nonplanar integrals, and includes all the nonplanar integrals at
two loops with five or less external legs.  In particular, all of the
nonplanar integrals in Figure~\ref{fig:fivePointAllDiags} are of this
type. For example diagram (a) can be made planar by moving 
external leg $3$.  Under this, the momenta of the propagators are modified
compared to the planar case at most by adding or subtracting a single
external momentum $p_k^{\mu}$.  Thus, the inverse
propagators ${\rho}_l$ therein can be written as either
$(x_i-x_j)^2$, or $(x_i-x_j\pm p_k)^2$, when using the dual
coordinates of the planar cousin.  The key observation here is that if
the infinitesimal boost vector $b^{\mu}$ is proportional to a massless
external leg $p_k^{\mu}$, then $(x_i-x_j\pm p_k)^2$ transforms in the
same way as $(x_i-x_j)^2$ for any $x_i^{\mu}$ and $x_j^{\mu}$.
Specifically,
\begin{equation}
\frac{\delta(x_i-x_j\pm p_k)^2}{(x_i-x_j\pm p_k)^2} =
\frac{\delta(x_i-x_j)^2}{(x_i-x_j)^2} =-b\cdot (x_i+x_j) \,,
\label{eq:keyIdentity}
\end{equation}
implying that all the
propagators in this class of nonplanar diagrams satisfy
$\delta {\rho}_k \propto {\rho}_k$ for this conformal
boost.

As a simple first example, consider the crossed double-box diagram on
the right of \fig{fig:fourPoints}, with numerator $N_1 = s u
(l_5+p_4)^2$, which is one of the nonplanar pure integrands found in
Ref.~\cite{Nimanonplanar} as a building block of the full amplitude:
\begin{equation} 
I^{\text{(np)}}=\int \mathcal{I}^{\text{(np)}} = 
\int d^D l_5 d^D l_6 \, \frac{N_1}{\prod_k  \rho_k}\,,
\end{equation}
where the $ \rho_k$ are the inverse propagators.
This diagram can be obtained from the planar double box in
\fig{fig:fourPoints} by moving  the external leg $3$ to the
central rung.  Using the dual coordinates of the planar double box, we
can write the nonplanar integrand as
\begin{equation}
 \mathcal{I}^{\text{(np)}} =
d^D x_5 d^D x_6 \, \frac{(x_1-x_3)^2 (x_2-x_1+p_3)^2
  (x_5-x_4)^2}{\prod_k  \rho_k} \,,
\end{equation}
where the propagators are given by
\begin{align}
{\rho}_1 = (x_5-x_1)^2,  & \hskip .4 cm {\rho}_2 = (x_5-x_2)^2,  \hskip .4 cm  {\rho}_3 = (x_5-x_3)^2,\nonumber \\ 
{\rho}_4 = (x_5-x_6)^2,  & \hskip .4 cm {\rho}_5  = (x_6-x_1)^2,  \hskip .4 cm {\rho}_6 = (x_6-x_4)^2, \nonumber \\
& \hskip - 2.45 cm {\rho}_7 = (x_5-x_6+p_3)^2,  &
\end{align}
with the $x_i$ defined in \eqn{TwoLoopDualVars}.
Applying a dual conformal transformation to the integrand with the boost vector $b^{\mu} \propto p_3^{\mu}$
and using equation (\ref{eq:keyIdentity}) we find that
\begin{equation}
\delta \mathcal{I}^{\text{(np)}}  = -(D-4)(b \cdot (x_5+x_6))\mathcal{I}^{\text{(np)}}\,,
\end{equation}
exposing a hidden symmetry in $D=4$.

A similar analysis holds for the numerator
$N_2=s t (l_5+p_3)^2$, corresponding to the other pure
integrand found in Ref.~\cite{Nimanonplanar}.  One can also obtain the crossed box from the planar double box by moving 
the leg $4$ to the central rung (and making a change in the momentum routing of the planar double box),
giving a new conformal boost with $b^{\mu} \propto p_4^{\mu}.$  As can be straightforwardly checked, both numerators $N_1$ and
$N_2$ give integrands that are
invariant in $D=4$ under this transformation as well.

While we propose these transformations as a natural extension of the
planar dual conformal symmetry, it is striking that the numerators
$N_{1}$ and $N_{2}$ are precisely the correct numerators of the
building blocks for the two-loop four-point amplitude in $\NeqFour$
sYM that unveil their analytic properties~\cite{Nimanonplanar}.  Here
we see that we can constrain these numerators from symmetry
considerations instead of from imposing desired analytic properties on
the integrands.  Similar symmetry considerations can be used to match
the numerators of a subset of three-loop four-point diagrams in
Ref.~\cite{nonplanarDlog} that can be obtained from planar ones by
moving a single external line. 

\begin{figure}[tb]
\includegraphics[scale=.11]{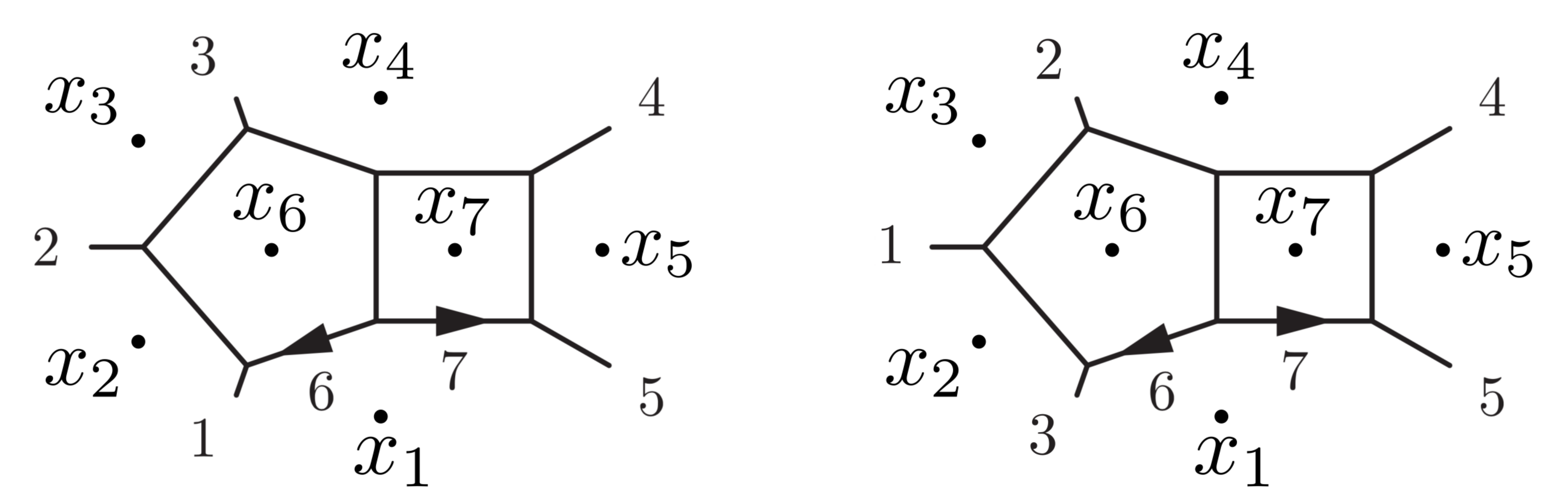}
\caption{Dual variables useful for the two-loop planar pentabox and the nonplanar integrals 
in \fig{fig:fivePointAllDiags}.}
\label{figLplanarEmbedding}
\end{figure}

\Section{Two-loop five-point case} 
As the central nontrivial example consider the two-loop five-point
$\NeqFour$ sYM amplitude first obtained in
Ref.~\cite{JJHenrikFivePointTwoLoopN4}.  This amplitude was rewritten
in a desired form where each diagram composing the amplitude contains
only logarithmic singularities and no pole at
infinity~\cite{nonplanarAmplituhedron}, as follows from dual conformal
symmetry in the planar case.  The diagrams composing this amplitude
are given in \fig{fig:fivePointAllDiags}.  These diagrams are either
planar, or in the nonplanar class of diagrams discussed above, so our
discussion immediately applies.

Consider diagrams (a), (d), (h), and (i), which can be
made planar by moving the external leg $3$,
corresponding to choosing $b^\mu \propto p^\mu_3$.  Using the dual coordinates
\begin{align}
p_1 &= x_3 - x_2\,,& p_2 &= x_4 - x_3\,, & p_3 &= x_2 - x_1\,, \nonumber\\
p_4 &= x_5 - x_4\,,& p_5 &= x_1 - x_5\,,& \nonumber\\
l_6 &= x_6-x_1 \,,& l_7 &= x_1-x_7 \,,&
\label{eq:TwoLoop_5pt_DualVars_1}
\end{align}
in the diagram on the right of Figure~\ref{figLplanarEmbedding}, the
propagators in the original nonplanar diagrams are a subset of
\begin{align}
&\rho_1 =(x_6-x_1)^2,     \hskip 1.35cm  \rho_2 = (x_6-x_3+p_3)^2 , \nonumber\\ 
&\rho_3 = (x_6-x_4+p_3)^2, \hskip .6 cm  \rho_4 = (x_7-x_4)^2, \nonumber\\ 
&\rho_5= (x_7-x_5)^2 ,   \hskip 1.4cm \rho_6 = (x_7-x_1)^2, \nonumber\\  
&\rho_7 = (x_6-x_7)^2,   \hskip 1.4cm \rho_8 = (x_6-x_7+p_3)^2.
\label{eq:fivePtProps}
\end{align}

A crucial difference between integrands at four points and five points
is the appearance of spinor helicity variables, which makes the transformation
properties less clear. We therefore restrict
to $D=4$ from now on, and the convention for spinors is chosen such
that 
$s_{ij}=(p_i+p_j)^2
 =\langle i j\rangle [ji] = \bra i j|i] = (\bra
  i_{\dot{a}}\ket{j}^{\dot{a}})([i|^{a} |j]_a)= \bra i p_j|i].$ A
complete set of numerators for the diagrams in
Figure~\ref{fig:fivePointAllDiags} is given in Table 3 of
Ref.~\cite{nonplanarAmplituhedron}.

To warm up, consider the numerator in diagram (i) 
\begin{equation}
N^{\rm (i)} = \bra{2} 4|3] \bra{3}5|2] - \bra{3} 4|2] \bra{2}5|3] \,.
\end{equation}
This numerator is constructed to follow the $S_3$ symmetry among legs $2,3,5$ of the diagram (up to a sign). By choosing to move leg $3$ to make the diagram planar and using the coordinates in Eq.~\eqref{eq:TwoLoop_5pt_DualVars_1}, we recast the numerator as
\begin{equation}
N^{\rm (i)} = \bra{3} x_{54} \, x_{43}\, x_{32}|3]+\bra{3} x_{23} \,x_{34} \, x_{45}|3]\,,
\label{eq:num_i_new}
\end{equation}
under momentum conservation and spinor identities. To see that this
numerator only rescales with a local weight under the transformation with $b^\mu \propto
p^\mu_3$, we need a nontrivial identity
\begin{equation}
	\frac{\delta \bra{b}x_{i_1i_2}\, x_{i_2i_3} \ldots x_{i_{n-1}i_n}|b]}
	{\bra{b}x_{i_1i_2} x_{i_2i_3} \ldots  x_{i_{n-1}i_n}|b]} 
	= -b\cdot(x_{i_1}+ \ldots + x_{i_n}) \,,
\label{eq:magicTrace}
\end{equation}
where $x_{ij}\equiv x_i-x_j$ and $\bra{b}x_{i_1i_2} \, x_{i_2i_3} \ldots x_{i_{n-1}i_n} |b]
 = (\bra{b}_{\dot{a}})(x^{\dot{a}a}_{i_1i_2})(x_{i_2i_3, a \dot{b}}) \ldots (x_{i_{n-1}i_n}^{\dot{c}d})(|b]_{d})$.   
We have confirmed Eq.~\eqref{eq:magicTrace} numerically through $n=8$,
irrespective of whether the $x_{ij}$'s are null separated or
not. Therefore the numerator in Eq.~\eqref{eq:num_i_new} is manifestly
rescaled under the transformation with weight $-b\cdot
(x_2+x_3+x_4+x_5)$.  Moreover, accounting for the transformation of
the propagators and measure using
Eqs.~\eqref{eq:deltaMeasure},~\eqref{eq:keyIdentity},
and~\eqref{eq:fivePtProps}, this is precisely the weight needed to
make the integrand invariant.

We can also make diagram (i) planar by  moving the leg
carrying momentum $p_2$ or $p_5$, giving a total of three choices of
$b^{\mu}$ for the conformal boosts.  We have checked that these three
transformations are independent symmetry generators, corresponding to
three hidden symmetries of this nonplanar integrand.

A more involved example is diagram (a) in
\fig{fig:fivePointAllDiags}. The numerator yielding the desired
analytic properties given in Ref.~\cite{nonplanarAmplituhedron} is
\begin{equation}
N_1^{\rm (a)} = 
\vev{13}\vev{24}\Big([24][13](l_7-l_7^*)^2 (l_6-l_6^*)^2 
  - (1\leftrightarrow 2) \Big)\,,
  \label{eq:numA_old}
\end{equation}
where $l_7^* = \frac{[54]}{[24]}\ket{5}[2|$ and 
$l_6^* =  p_1+\frac{[23]}{[13]}|2\rangle [1|$.
 How this numerator transforms is far from clear in the
  above form.  In fact, the first or second term alone does not
  rescale with a local weight.  However, by using on-shell conditions and Schouten
  identities it can be rewritten as
\begin{align}
N_1^{\rm (a)} &= -\bra{3}x_{23}\, x_{34} \, x_{45} |3]\rho_4\rho_1 \label{NumeratorA}  \\
&\quad +\bra{3}x_{23} \, x_{34}\, x_{45}\, x_{57}\, x_{76}\, x_{61}\, x_{14}|3] \,, \nonumber 
\end{align}
using the dual coordinates in Eq.~\eqref{eq:TwoLoop_5pt_DualVars_1}.
With the help of equation (\ref{eq:magicTrace}), each of the two terms in
\eqn{NumeratorA} above transform with the weight necessary to
make the integrand invariant in $D=4$. 
After canceling the propagators, the first term gives rise to the daughter diagram 
(i) in \fig{fig:fivePointAllDiags}, and the numerator $\bra{3}x_{23}x_{34}x_{45}|3]$
also matches to one of the components in Eq.~\eqref{eq:num_i_new}.

Similarly, we can rewrite the original numerators of diagrams (d) and
(h) using the dual coordinates in the diagram on the left of
Fig.~\ref{figLplanarEmbedding} as
\begin{align}
N_1^{\rm (d)} 
&= s_{34}(s_{34}+s_{35}) \Bigl(l_7-\frac{\langle 54\rangle}{\langle 34\rangle} |3\rangle [5| \Bigr)^2 \nonumber \\
&= s_{34}(s_{34}+s_{35}) \rho_6 + \bra{3} x_{71} \,  x_{15} \, x_{54} |3]\,,
\end{align}
and
\begin{align}
N_1^{\rm (h)} 
&= \langle 15\rangle [35] \langle 23\rangle [12] \Bigl(l_6-\frac{\langle 12\rangle}{\langle 32\rangle} |3\rangle [1| \Bigr)^2 \nonumber \\
&= (s_{23}s_{35} - \bra{3} x_{34}\, x_{45}\, x_{51}|3]) \rho_1 \nonumber \\
&  \null \hskip 1.8 cm 
  - s_{12}\bra{3} x_{62}\, x_{23}\ x_{35} |3] \,, \hskip .3 cm \nonumber \\
N_3^{\rm (h)} 
&= -s_{12} \bra{3}p_1 p_5 l_6 |3] = -s_{12} \bra{3} x_{35}\, x_{51} \, x_{16} |3] \,.  \hskip .1 cm 
\end{align}
In addition there are numerators simply related via diagram symmetries.
Using Eqs.~\eqref{eq:deltaMeasure},~\eqref{eq:keyIdentity},
and~\eqref{eq:magicTrace}, we see that these numerators have weights
that make the integrand invariant under the dual conformal boost
with $b^\mu \propto p^\mu_3$.

Diagrams (c) and (f) can be made planar by moving the external leg
carrying momentum $p_4$, corresponding to $b^\mu \propto p^\mu_4$. 
The dual coordinates are defined according to the left of
Figure~\ref{figLplanarEmbedding}, analogous to Eq.~\eqref{eq:TwoLoop_5pt_DualVars_1}.
The propagators in the original
nonplanar diagrams are a subset of
\begin{align}
\rho_1 = (x_6-x_1)^2, & \hskip .4 cm \rho_2 = (x_6-x_2)^2, & \hskip -.8 cm \rho_3 = (x_6-x_3)^2,   \nonumber \\ 
\rho_4 = (x_6-x_4)^2, & \hskip .4 cm \rho_5 = (x_7-x_5)^2, & \hskip -.8 cm \rho_6 = (x_7-x_1)^2, \nonumber \\  
\rho_7 = (x_6-x_7)^2, & \hskip .4 cm \rho_8 = (x_6-x_7+p_4)^2.  &  
\end{align}
The numerator of diagram (f) is $N_1^{\rm (f)}= s_{14}s_{45}
(l_6+p_5)^2$ which manifestly rescales with local weight under the
transformation.  To see the conformal property of diagram (c), we need 
\begin{align}
N_1^{\rm (c)} 
&= \langle 15\rangle [54] \langle 43\rangle [13] \left(l_6-l^*_6\right)^2 (l_6+p_4)^2 \\
&= 
\left( -s_{51} s_{45} \rho_3 + \bra{4}x_{46}\, x_{63} \, x_{32}\, x_{21} \, x_{15} |4] \right) (l_6+p_4)^2, \nonumber 
\end{align}
with the same $l^*_6$ as defined below Eq.~\eqref{eq:numA_old}.
After canceling the propagator, the first term matches $N_2^{\rm (f)}$
of Ref.~\cite{nonplanarAmplituhedron} which is related to $N_1^{\rm
  (f)}$ under $4\leftrightarrow 5$.

We have checked all of the two-loop five-point nonplanar
integrands from Ref.~\cite{nonplanarAmplituhedron} that manifest the
desired analytic properties of the full two-loop five-point amplitude
and found that all of them have a hidden symmetry in $D=4$ closely
related to dual conformal symmetry.  In cases where more than one
conformal boost is available, as for diagrams (c), (f), (h), and (i)
in Figure~\ref{fig:fivePointAllDiags}, we have checked that all
such choices of $b^{\mu}$ give symmetries of the integrand.  While
\eqn{eq:keyIdentity} guarantees that all the propagators transform
with definite weight, the fact that all the corresponding numerators
behave accordingly to make the integrand invariant appears miraculous.

Using equations (\ref{eq:keyIdentity}) and (\ref{eq:magicTrace}) we
can generalize these results to integrals relevant for higher-point amplitudes.  As a concrete example,
consider diagram (a) in Figure~\ref{fig:fivePointAllDiags} but with
legs 1,2,4,5 being massive or replaced with arbitrary
collections of massless particles, while keeping leg 3 massless.
Crucially, the identity in Eq.~\eqref{eq:magicTrace} holds even for
$x^2_{i,i+1} \neq 0$. This implies the numerator with the dual
variables in Eq.~\eqref{eq:TwoLoop_5pt_DualVars_1}
\begin{equation}
\bra{3}x_{23}\,x_{34}\,x_{45}\,x_{57}\,x_{76}\,x_{61}\,x_{14}|3]\,,
\end{equation}
transforms with the proper weight to make the integrand
invariant, providing a generalization of the second term in \eqn{NumeratorA}. Another possible numerator is
\begin{equation}
\label{eq:finiteNumerator}
s_{12} s_{24} \big ( \bra{3} x_{47}\,x_{76}\,x_{61}|3] + \bra{3} x_{16}\,x_{67}\,x_{74} |3] \big ) \, .
\end{equation}
The latter example~\eqref{eq:finiteNumerator} is especially
interesting since it vanishes in the collinear limit $x_{76}^\mu \propto
p_3^\mu$ and gives an infrared-finite integral, for which the hidden
symmetry is \emph{exact} and free of anomalies from divergences. By
working in six dimensions, additional finite integrals with the hidden
symmetry can be found; such integrals are related to four-dimensional
ones via dimension shifting relations~\cite{DimensionShifting}.

\Section{Conclusions}
Following the four-point hints in Ref.~\cite{EarliernonplanarSym},
here we demonstrated that all sectors of the two-loop five-point
$\NeqFour$ sYM amplitude, including the nonplanar sector, possess new
nontrivial hidden symmetries related to dual conformal symmetry.  To
show this we demonstrated that each integrand sector identified in
Ref.~\cite{nonplanarAmplituhedron} possessing simple analytic
properties manifests a hidden symmetry.  For some sectors the symmetry
is rather unobvious. The construction used for the two-loop five-point
amplitude extends to any number of loops and legs, giving an infinite
class of integrands with new hidden symmetries.  It would be
interesting to check if these cases actually appear with nonzero
coefficient in $\NeqFour$ sYM amplitudes.  Even for the cases studied
here we can expect a larger set of symmetries than the ones we found;
we expect this to be helpful for the important problem of identifying
the hidden symmetries of more general cases beyond the ones studied
here.  It would be interesting to apply the symmetries to help
identify nonplanar integrals of uniform transcendentality, which
become nontrivial at high loop orders by directly checking leading
singularities~\cite{IRN4}.  It would also be interesting to understand
how the new symmetries described here relate to recent progress in
extending integrability to nonplanar theories described in
Ref.~\cite{nonplanarWilson}.  Given the useful role hidden symmetries
have played in the planar sector of $\NeqFour$ sYM theory, we should
expect new progress from fully unraveling the corresponding
symmetries of the non-planar sector of the theory.

\vskip .2 cm 
\noindent {\it Acknowledgments.}
We thank Lance Dixon, Enrico Herrmann, Harald Ita, Julio Parra-Martinez, and Jaroslav Trnka for discussions. This work was
supported in part by the Department of Energy under Award Number DE-SC0009937.


\begin{thebibliography}{99}

\bibitem{DualConformal}
J.~M.~Drummond, J.~Henn, V.~A.~Smirnov and E.~Sokatchev,
JHEP {\bf 0701}, 064 (2007)
[hep-th/0607160];\\
%
Z.~Bern, M.~Czakon, L.~J.~Dixon, D.~A.~Kosower and V.~A.~Smirnov,
Phys.\ Rev.\ D {\bf 75}, 085010 (2007)
  [hep-th/0610248];\\
%
J.~M.~Drummond, G.~P.~Korchemsky and E.~Sokatchev,
Nucl.\ Phys.\ B {\bf 795}, 385 (2008)
[arXiv:0707.0243 [hep-th]];\\
%
J.~M.~Drummond, J.~Henn, G.~P.~Korchemsky and E.~Sokatchev,
Nucl.\ Phys.\ B {\bf 828}, 317 (2010)
[arXiv:0807.1095 [hep-th]].

\bibitem{Yangian}
I.~Bena, J.~Polchinski and R.~Roiban,
Phys.\ Rev.\ D {\bf 69}, 046002 (2004)
[hep-th/0305116];\\
%
J.~M.~Drummond, J.~M.~Henn and J.~Plefka,
JHEP {\bf 0905}, 046 (2009)
[arXiv:0902.2987 [hep-th]];\\
%
N.~Beisert, J.~Broedel and M.~Rosso,
J.\ Phys.\ A {\bf 47}, 365402 (2014)
[arXiv:1401.7274 [hep-th]].

\bibitem{BeisertStaudacher}
N.~Beisert and M.~Staudacher,
Nucl.\ Phys.\ B {\bf 670}, 439 (2003)
[hep-th/0307042].

\bibitem{WilsonLoops}
J.~M.~Drummond, J.~Henn, G.~P.~Korchemsky and E.~Sokatchev,
Nucl.\ Phys.\ B {\bf 815}, 142 (2009)
[arXiv:0803.1466 [hep-th]];\\
%
A.~V.~Belitsky, G.~P.~Korchemsky and E.~Sokatchev,
Nucl.\ Phys.\ B {\bf 855}, 333 (2012)
[arXiv:1103.3008 [hep-th]];\\
%
L.~J.~Mason and D.~Skinner,
JHEP {\bf 1012}, 018 (2010)
[arXiv:1009.2225 [hep-th]].

\bibitem{UniformTransc}
A.~V.~Kotikov, L.~N.~Lipatov, A.~I.~Onishchenko and V.~N.~Velizhanin,
Phys.\ Lett.\ B {\bf 595}, 521 (2004)
Erratum: [Phys.\ Lett.\ B {\bf 632}, 754 (2006)]
 [hep-th/0404092];\\
%
A.~V.~Kotikov and L.~N.~Lipatov,
Nucl.\ Phys.\ B {\bf 769}, 217 (2007)
[hep-th/0611204].

\bibitem{BassoOPE}
B.~Basso, A.~Sever and P.~Vieira,
Phys.\ Rev.\ Lett.\  {\bf 111}, no. 9, 091602 (2013)
[arXiv:1303.1396 [hep-th]];\\
%
 B.~Basso, A.~Sever and P.~Vieira,
  JHEP {\bf 1401}, 008 (2014)
  [arXiv:1306.2058 [hep-th]].

\bibitem{DixonBootstrap}
L.~J.~Dixon, J.~M.~Drummond, C.~Duhr and J.~Pennington,
JHEP {\bf 1406}, 116 (2014)
[arXiv:1402.3300 [hep-th]];\\
%
L.~J.~Dixon and M.~von Hippel,
JHEP {\bf 1410}, 065 (2014)
[arXiv:1408.1505 [hep-th]];\\
%
L.~J.~Dixon, J.~Drummond, T.~Harrington, A.~J.~McLeod, G.~Papathanasiou and M.~Spradlin,
JHEP {\bf 1702}, 137 (2017)
[arXiv:1612.08976 [hep-th]];\\
%
S.~Caron-Huot, L.~J.~Dixon, A.~McLeod and M.~von Hippel,
Phys.\ Rev.\ Lett.\  {\bf 117}, no. 24, 241601 (2016)
[arXiv:1609.00669 [hep-th]].

\bibitem{BDS}
Z.~Bern, L.~J.~Dixon and V.~A.~Smirnov,
Phys.\ Rev.\ D {\bf 72}, 085001 (2005)
[hep-th/0505205];\\
%
L.~F.~Alday and J.~M.~Maldacena,
JHEP {\bf 0706}, 064 (2007)
[arXiv:0705.0303 [hep-th]].

\bibitem{Grassmannian}
N.~Arkani-Hamed, J.~L.~Bourjaily, F.~Cachazo, A.~B.~Goncharov, A.~Postnikov and J.~Trnka,
arXiv:1212.5605 [hep-th].

\bibitem{Amplituhedron}
N.~Arkani-Hamed and J.~Trnka,
JHEP {\bf 1410}, 030 (2014)
[arXiv:1312.2007 [hep-th]];\\
%
N.~Arkani-Hamed and J.~Trnka,
JHEP {\bf 1412}, 182 (2014)
[arXiv:1312.7878 [hep-th]].

\bibitem{Symbols}
A.~B.~Goncharov, M.~Spradlin, C.~Vergu and A.~Volovich,
Phys.\ Rev.\ Lett.\  {\bf 105}, 151605 (2010)
[arXiv:1006.5703 [hep-th]].

\bibitem{HennIntegrals}
J.~M.~Henn,
Phys.\ Rev.\ Lett.\  {\bf 110}, 251601 (2013)
[arXiv:1304.1806 [hep-th]].

\bibitem{nonplanarDlog}
Z.~Bern, E.~Herrmann, S.~Litsey, J.~Stankowicz and J.~Trnka,
JHEP {\bf 1506}, 202 (2015)
[arXiv:1412.8584 [hep-th]].

\bibitem{nonplanarAmplituhedron}
Z.~Bern, E.~Herrmann, S.~Litsey, J.~Stankowicz and J.~Trnka,
JHEP {\bf 1606}, 098 (2016)
[arXiv:1512.08591 [hep-th]].

\bibitem{JJHenrikFivePointTwoLoopN4}
J.~J.~Carrasco and H.~Johansson,
Phys.\ Rev.\ D {\bf 85}, 025006 (2012)
[arXiv:1106.4711 [hep-th]].

\bibitem{HennFivePtTwoloop}
T.~Gehrmann, J.~M.~Henn and N.~A.~Lo Presti,
Phys.\ Rev.\ Lett.\  {\bf 116}, no. 6, 062001 (2016)
Erratum: [Phys.\ Rev.\ Lett.\  {\bf 116}, no. 18, 189903 (2016)]
[arXiv:1511.05409 [hep-ph]];\\
%
D.~Chicherin, J.~Henn and V.~Mitev,
JHEP {\bf 1805}, 164 (2018)
[arXiv:1712.09610 [hep-th]].

\bibitem{NimaAllLoopIntegrand}
N.~Arkani-Hamed, J.~L.~Bourjaily, F.~Cachazo, S.~Caron-Huot and J.~Trnka,
JHEP {\bf 1101}, 041 (2011)
[arXiv:1008.2958 [hep-th]].

\bibitem{ParkeTaylor}
S.~J.~Parke and T.~R.~Taylor,
Phys.\ Rev.\ Lett.\  {\bf 56}, 2459 (1986);\\
%
M.~L.~Mangano, S.~J.~Parke and Z.~Xu,
Nucl.\ Phys.\ B {\bf 298}, 653 (1988).

\bibitem{Nimanonplanar}
N.~Arkani-Hamed, J.~L.~Bourjaily, F.~Cachazo and J.~Trnka,
Phys.\ Rev.\ Lett.\  {\bf 113}, no. 26, 261603 (2014)
[arXiv:1410.0354 [hep-th]].

\bibitem{nonplanarParkeTaylor}
N.~Arkani-Hamed, J.~L.~Bourjaily, F.~Cachazo, A.~Postnikov and J.~Trnka,
JHEP {\bf 1506}, 179 (2015)
[arXiv:1412.8475 [hep-th]].

\bibitem{nonPlanarOnShellDiagrams}
S.~Franco, D.~Galloni, B.~Penante and C.~Wen,
JHEP {\bf 1506}, 199 (2015)
[arXiv:1502.02034 [hep-th]];\\
%
J.~L.~Bourjaily, S.~Franco, D.~Galloni and C.~Wen,
JHEP {\bf 1610}, 003 (2016)
[arXiv:1607.01781 [hep-th]].

\bibitem{NonplanarYangian}
R.~Frassek and D.~Meidinger,
JHEP {\bf 1605}, 110 (2016)
[arXiv:1603.00088 [hep-th]].

\bibitem{EarliernonplanarSym}
Z.~Bern, M.~Enciso, H.~Ita and M.~Zeng,
Phys.\ Rev.\ D {\bf 96}, no. 9, 096017 (2017)
[arXiv:1709.06055 [hep-th]].

\bibitem{DENoDoubledProp}
  M.~Zeng,
  JHEP {\bf 1706}, 121 (2017)
  [arXiv:1702.02355 [hep-th]];\\
  %
  J.~Bosma, K.~J.~Larsen and Y.~Zhang,
  Phys.\ Rev.\ D {\bf 97}, no. 10, 105014 (2018)
  [arXiv:1712.03760 [hep-th]].

\bibitem{DimensionShifting}
 Z.~Bern, L.~J.~Dixon and D.~A.~Kosower,
  Phys.\ Lett.\ B {\bf 302}, 299 (1993)
  Erratum: [Phys.\ Lett.\ B {\bf 318}, 649 (1993)]
  [hep-ph/9212308]; 
%
  Nucl.\ Phys.\ B {\bf 412}, 751 (1994)
  [hep-ph/9306240];\\
%
O.~V.~Tarasov,
Phys.\ Rev.\ D {\bf 54}, 6479 (1996)
[hep-th/9606018].

\bibitem{IRN4}
R.~H.~Boels, T.~Huber and G.~Yang,
JHEP {\bf 1801}, 153 (2018)
[arXiv:1711.08449 [hep-th]].

\bibitem{nonplanarWilson}
B.~Eden, Y.~Jiang, D.~le Plat and A.~Sfondrini,
JHEP {\bf 1802}, 170 (2018)
[arXiv:1710.10212 [hep-th]];\\
T.~Bargheer, J.~Caetano, T.~Fleury, S.~Komatsu and P.~Vieira,
arXiv:1711.05326 [hep-th];\\
R.~Ben-Israel, A.~G.~Tumanov and A.~Sever,
arXiv:1802.09395 [hep-th].

\end{thebibliography}
\end{document}